# Time-evolution of the GaAs(0 0 1) pre-roughening process

Z. Ding [a], D.W. Bullock [a], P.M. Thibado [a], V.P. LaBella [b], and Kieran Mullen [c]

[a] Department of Physics, The University of Arkansas, Physics Building, Room 220, Fayetteville, AR 72701, USA
[b] School of NanoSciences and NanoEngineering, University at Albany-SUNY, Albany, NY 12203, USA
[c] Department of Physics and Astronomy, The University of Oklahoma, Norman, OK 73019, USA

Abstract
The GaAs(0 0 1) surface is observed to evolve from being perfectly flat to a surface half covered with one-monolayer high spontaneously formed GaAs islands. The dynamics of this process are monitored with atomic-scale resolution using scanning tunneling microscopy. Surprisingly, pit formation dominates the early stages of island formation. Insight into the nucleation process is reported.

1. Introduction

Numerous electronic devices have been fabricated from compound semiconductor materials, such as GaAs, grown by epitaxial methods. Epitaxy involves depositing atoms or molecules in an atomic layer-by-layer fashion onto a clean, atomically flat single crystal surface. Probing the dynamics of how adatoms interact with these surfaces is both fundamentally and technologically important.

One successful approach for understanding the dynamics is to monitor a surface evolving toward equilibrium on the atomic scale. On single component metallic surfaces much has been learned about the atomic-scale kinetics and dynamics by watching how deposited islands decay in time [1–6]. In these studies a sub-monolayer amount of material is added to the surface in the form of islands. In time, the atoms leave the islands and attach themselves to terrace edges (present due to a slight miscut) to lower the total number of undercoordinated atoms. These studies are modeled by considering the minimization of the internal energy of the system (i.e., Ostwald ripening) [7].

The kinetics and dynamics of two components compound semiconductor surfaces are more challenging, and few atomic-scale investigations have been carried out. In the experiment by Johnson et al. the authors observed atomic-scale roughening on GaAs(0 0 1) during growth [8]. From this, Tersoff, Johnson and Orr concluded that the surface is very close to equilibrium at normal growth conditions and entropy considerations are important [9]. Recently, the equilibrium thermodynamic properties of the GaAs(0 0 1) surface were successfully modeled as a 2D Ising system [10]. From this study it was found that islands spontaneously form when the GaAs(0 0 1) surface is heated and no material is deposited. The non-equilibrium dynamics of the spontaneous island formation process, however, have not been studied.

In this article, scanning tunneling microscopy (STM) is used to image the far from equilibrium dynamics and atomic-scale configurations of a GaAs(0 0 1) surface as it relaxes toward an equilibrium configuration specifically chosen as a surface half covered with GaAs islands. The time-resolved STM images of the domain structures (i.e., islands) on the GaAs(0 0 1) surface show that the system passes through an intermediate pit formation stage as it goes from a fully unoccupied or flat surface to a half occupied surface.

2. Experimental

Experiments were carried out in an ultrahigh vacuum (UHV) multi-chamber facility ($5-8 \times 10^{-11}$ Torr throughout) that contains a solid-source molecular beam epitaxy (MBE) chamber (Riber 32P) with a substrate temperature determination system accurate to ±2 C [11], and an arsenic cell with an automated valve and controller. The MBE chamber also has an all UHV connection to a surface analysis chamber, which contains a custom integrated STM (Omicron) [12]. Commercially available "epi-ready," n+ (Si doped $10^{18}/cm^3$) GaAs(0 0 1) ±0.01 degree substrates were loaded into the MBE system without any chemical cleaning. The surface oxide layer was removed and a 1.5-μm thick GaAs buffer layer was grown at 580 C using an $As_4$ to Ga beam equivalent pressure (BEP) ratio of 15 and a growth rate of 1.0 μm/h as determined by reflection high energy electron diffraction (RHEED) oscillations.

To study the non-equilibrium dynamics of the spontaneously formed islands, a procedure was developed using the known equilibrium properties of the system [10]. First, the substrate was annealed at a temperature above which pre-roughening occurs (527.5 C), while exposing it to an $As_4$ BEP that produced a flat, island free surface. Next, in less than a minute, the $As_4$ BEP was lowered to a level that produces a surface half covered with islands once equilibrated. Reaching equilibrium would require 2–16 h of annealing, but the substrates were quenched to room temperature for various annealing times shorter than this to produce a time series of samples for each temperature. The quenching algorithm used freezes in the surface morphology present at the higher temperatures and has been described elsewhere [13]. The samples were transferred to the STM without breaking UHV and imaged at room temperature. All samples prepared at different times were prepared by first repeating the entire procedure starting with a regrowth of the buffer layer. This produced several series of GaAs(0 0 1) surfaces with the same initial conditions and increasing anneal times.

For every sample, 5–10 1 μm×1 μm filled-state STM images were acquired using tips made from single crystal <111>-oriented tungsten wire, a sample bias of -3.0 V, and a tunneling current of 0.05–0.1 nA. To compute the fractional coverage of the surface covered by the islands, 10–20 200 nm × 200 nm regions are cropped far from terrace edges from 5–10 larger images, and then thresholded to compute an average coverage, which has a uniform standard deviation of ~5%.

3. Results

The time-evolution of the nanometer scale configuration of the spontaneously formed islands for the GaAs(0 0 1) surface is visually displayed in a series of 1 μm×1 μm STM images in Fig. 1. The images show the surface after being annealed for successively longer times as indicated, while the substrate was held at 560 C and while the surface was exposed to a 0.03 μTorr $As_4$ BEP. Initially ($t$ = 0), the surface is flat, nearly free of islands or pits. After only a 5 min anneal, one-monolayer high islands and one-monolayer deep pits pepper the surface. Notice that the pits are larger than the islands, however the number density of islands is greater than the number density of pits. In addition, the islands and pits are elongated in the [1-10] direction and double-height steps are never formed. As the time increases, the islands and pits grow in size. Interestingly, the pits eventually disappear and the surface is half covered with islands and no change in the morphology is observed for longer times indicating that equilibrium has been reached [10,14]. The STM images of the GaAs(0 0 1) surface annealed at 570 _C are shown in Fig. 2. Surprisingly, the spontaneously formed islands show the same time-evolution behavior as those at the lower temperature shown in Fig. 1. However, notice that it takes 30 min for sample at 560 C to reach the specific morphology, while it takes only 15 min to reach the similar morphology for sample at 570 C. In fact, at every stage of the relaxation process the time at 570 C is half the time required at 560 C.

To further quantify the time-evolution properties, a decay time constant was found for each substrate annealing temperature studied. The various decay time constants, $\tau$ are plotted versus the substrate temperature in Fig. 3. Notice, as expected, that the time to reach equilibrium becomes longer as the temperature is decreased. A linear fit yields a large activation energy of 4.2 eV.

4. Discussion

Observations of islands spontaneously forming on the GaAs(0 0 1) surface have been previously reported and shown to arise from entropic factors controlling the free energy minimization process [10]. The first 5 min of the spontaneous island formation process tells a surprising story about the pathway to equilibrium. Mainly, that the Ga atoms which form the GaAs islands come from the middle of the terraces via pit formation. This observation is contrary to the typical assumption that atoms come from the terrace edges, since atoms at the terrace edges are under-coordinated, requiring less energy to break free when compared to an atom embedded in the middle of the terrace. However, this general policy assumes there is a coordination difference between these two atomic sites, which is not true for the GaAs(001)-β2(2×4) surface. The atomic structure model for the GaAs(001)-β2(2×4) reconstructed surface is locally multi-leveled or corrugated such that every unit cell on the pristine terrace is nearly identical to the local structure at a terrace edge site, as shown in Fig. 4 [13]. Given that there is a much larger number of middle terrace sites compared to edge terrace sites, it is clear why pit formation is preferred.

In the beginning stages of island formation the coverage of pits and islands are equal. This indicates that the islands were formed by making adjacent pits. In time, however, the pits preferentially spread out in the [1-10] direction and merge into each other forming larger pits. This is the direction of the top-layer arsenic dimer bond, and has been previously shown to be the fast diffusion and preferential growth direction [15]. Here, we see it is also the preferential etching or dissociation direction. Once the pits grow to the size of the terrace, they punch through the edge of the terrace forming an ''inlet-'' or ''bay-type'' structure. The geometry of the inlet structure slowly evolves from something long and narrow to no ''inlet'' at all. The islands on the other hand do not grow in size in the beginning. The small islands tend to be very stable and kinetics drives the formation of more islands rather than making the existing islands bigger (even though this is what the pits prefer to do). In addition, double-height steps never form, so the islands tend to stay small to avoid the steps at the edges of the pits.

Once a small pit is formed, atoms continue to break away from the edge of the pit and climb out of the pit to form more islands on the terrace. This indicates that there is no difference between the energy barrier for an atom to diffuse over a step and an atom to diffuse across a flat terrace (i.e., no significant Schwoebel barrier) [16]. The lack of a Schwoebel barrier on the GaAs(001)-$\beta2(2\times4)$ reconstructed surface is simply due to the same atomic structure existing at the edge of the terrace and the middle of the terrace as shown in Fig. 4, i.e., diffusion across a pristine terrace is identical to diffusion across a terrace edge or step. This is interesting because without a Schwoebel barrier, we know individual atoms can easily flow over step edges. However, small islands must also avoid the step edge, thus there is a many-body requirement playing a significant role in the dynamics. This higher-order effect must be considered when modeling the surface kinetics using atomistic models.

It is interesting to notice that the pits eventually disappear with the time. A large terrace width helps the investigation because it minimizes any effects due to terrace edges, however the large terrace is responsible for creating two time scales in this experiment. The first one is the time to reach the equilibrium coverage, while the second time is a measure of how long it takes for the pits to diffuse to the terrace edge. Finally, the favoring of islands over pits indicates that inside corners cost more energy than outside corners [14].

Preferential pit formation also gives insight into why epitaxy is relatively easy on GaAs(0 0 1). Deposited atoms do not need to travel very far to nucleate since the reconstruction has steps built into it. In fact, all homoepitaxy may be easier on step-containing surface reconstructions (e.g., easier on GaAs(0 0 1) versus GaAs(1 1 0)).

A quantitative analysis of how the time to equilibrium changes with temperature is necessary to understand the physics of the data shown in Fig. 3. Clearly, the time gets longer as the temperature is lowered. If we model the temperature dependence using an Arrhenius relationship we get an activation energy of 4.2 eV [8,9]. However, the value is larger than the energy required to evaporate Ga from GaAs(0 0 1) [17]. It is also four times larger than both the experimentally measured and first principles calculated activation energy for Ga diffusion (1.2 eV) [18,19]. This suggests the phenomenon is slowing down much faster than a thermally activated process. One way to predict such a rapid decrease in kinetics is to speculate that the surface is undergoing a phase transition. For critical phenomenon, the decay time constant should diverge as the critical temperature is approached [20–22]. This is consistent with the earlier findings that this surface follows the equilibrium properties of a 2D Ising system. Proving this system is experiencing a critical slowing down would shed light on whether or not phenomenological kinetic modeling is appropriate for the GaAs(0 0 1) surface.

5. Conclusion

In summary, atomic-scale, time-resolved STM images of the dynamics of the spontaneous island formation process on the GaAs(0 0 1) surface was presented. The early stages of island formation are controlled by pit formation. This proves that atoms preferentially break away from the middle of the terrace versus the edge of the terrace as previously thought. The atomic structural model for the GaAs surface provides an explanation for why atoms are more likely to break away for the terrace and why epitaxy is more successful on the (0 0 1) surface.

Acknowledgements

Special thanks to R. Hauenstein for his helpful comments. This work is supported by the National Science Foundation grant FRG DMR-0102755 and the State of New York grant NYSTAR-FDPC020095.

Fig. 1. Series of 1 µm×1 µm STM images of the GaAs(001) surface showing the time-evolution of the spontaneously formed islands while holding the substrate temperature at 560 C. The morphology changes from flat terraces, to one with islands and pits, and finally to flat terraces with islands that cover 50% of the surface. All images are displayed using a grayscale with each shade representing a 1 ML (0.3 nm) height change.

Fig. 2. Series of 1 µm×1 µm STM images of the GaAs(001) surface showing the time-evolution of the spontaneously formed islands while holding the substrate temperature at 570 C. All images are displayed using a gray-scale with each shade representing a 1 ML (0.3 nm) height change.

Fig. 3. Decay time constant in seconds for reaching equilibrium versus substrate anneal temperature (data points). Linear fit is shown as a solid line.

Fig. 4. Top view and side view of the GaAs(001)-β2(2×4) structural model. The right side of the model is a flat terrace, while the left side shows a single step up that would occur at the edge of the terrace. Notice, the edge of the terrace has the same structure that is present within the flat terrace.

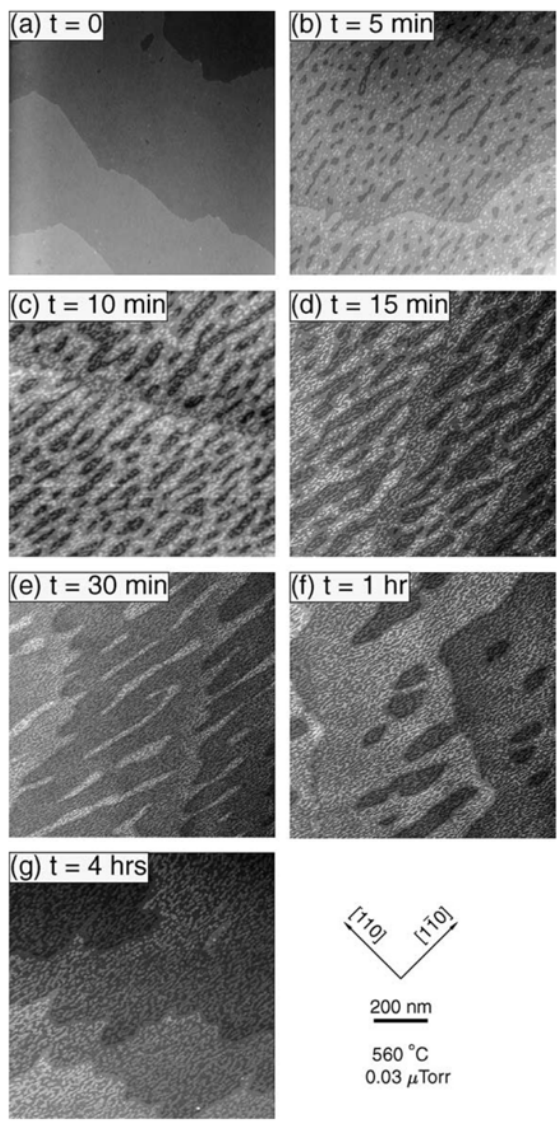

Figure 1.

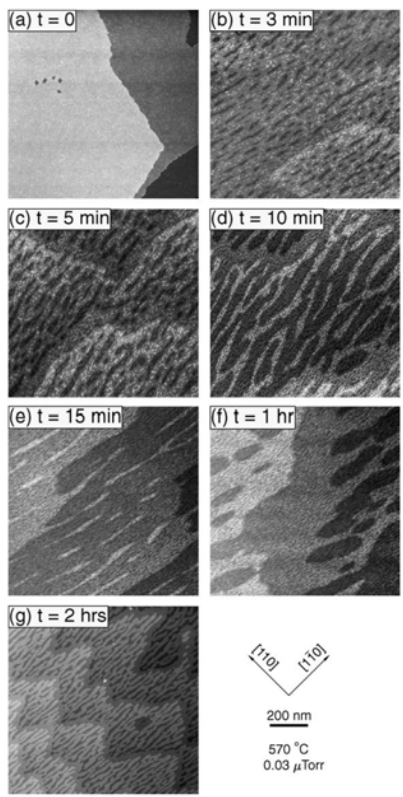

Figure 2.

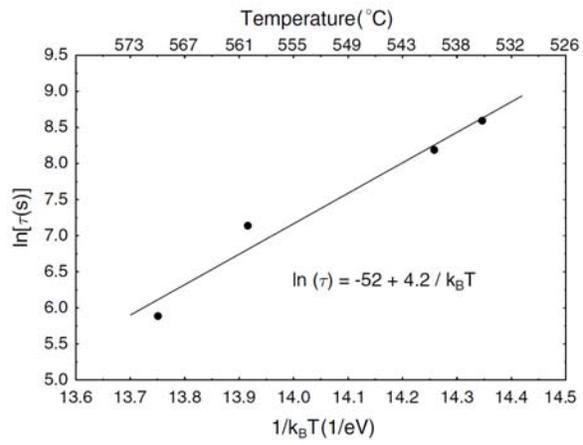

Figure 3.

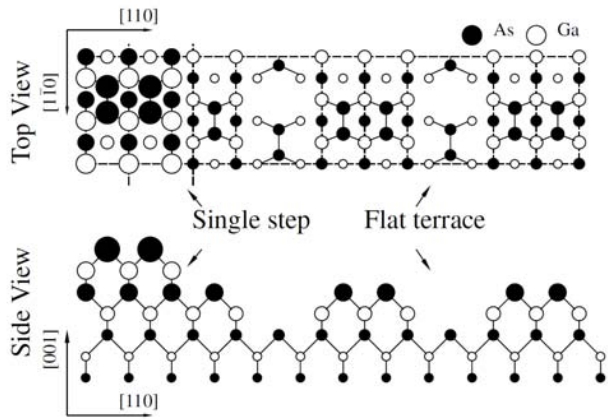

Figure 4.